\documentclass[amsmath,aps,floats,amssymb, floatfix, superscriptaddress, twocolumn,nofootinbib]{revtex4}

\usepackage{graphicx}
\usepackage{amsmath}
\usepackage{color}

\def\ltwid{\mathrel{\raise.3ex\hbox{$<$\kern-.75em\lower1ex\hbox{$\sim$}}}}
\def\gtwid{\mathrel{\raise.3ex\hbox{$>$\kern-.75em\lower1ex\hbox{$\sim$}}}}

\def\square{\kern1pt\vbox{\hrule height 1.2pt\hbox{\vrule width 1.2pt\hskip 3pt
   \vbox{\vskip 6pt}\hskip 3pt\vrule width 0.6pt}\hrule height 0.6pt}\kern1pt}
\def\overleftrightarrow#1{\vbox{\ialign{##\crcr
     $\leftrightarrow$\crcr\noalign{\kern-1pt\nointerlineskip}
     $\hfil\displaystyle{#1}\hfil$\crcr}}}

\def\zetaa{\zeta_{1}}
\def\zetaaa{\zeta_{2}}

\newcommand{\be}{\begin{equation}}
\newcommand{\ee}{\end{equation}}
\newcommand{\bea}{\begin{eqnarray}}
\newcommand{\eea}{\end{eqnarray}}
\newcommand{\nn}{\nonumber}

\newcommand{\Ec}[1]{(\ref{eq:#1})}

\newcommand{\eql}[1]{\label{eq:#1}}

\begin{document}

\title{Growth of perturbations in nonlocal gravity with non-$\Lambda$CDM background}

\author{Sohyun Park}
\email{spark1@kasi.re.kr}
\affiliation{Korea Astronomy and Space Science Institute,
Daejeon, 305-348, Korea}

\author{Arman Shafieloo}
\email{shafieloo@kasi.re.kr}
\affiliation{Korea Astronomy and Space Science Institute,
Daejeon, 305-348, Korea}
\affiliation{University of Science and Technology, Daejeon 34113, Korea}

\begin{abstract}
\noindent
We re-analyze the nonlocal gravity model of Deser and Woodard which was proposed to account for the current phase of cosmic acceleration. We show that the growth of perturbations predicted by this nonlocal gravity model when its background evolution is fixed by some particular non-$\Lambda$CDM models can be substantially lower than when its background is fixed by $\Lambda$CDM. 
This can be seen when we consider the background expansion by a dark energy model with a slightly less negative equation of state with respect to cosmological constant. 
Our results hints towards a fact that the choice of the background expansion can play a crucial role how this nonlocal gravity model can fit the growth history data. 
While the growth data might show better consistency to GR models (among the background models we studied so far), it seems the nonlocal gravity model studied in this work is able to show comparable consistency to the growth data as well. Showing this consistency can be considered as a significant result since this model can stand as a distinguishable alternative to the standard model of cosmology.
\end{abstract}


\maketitle

\section{Introduction}\label{intro}

Evidences for the acceleration of cosmic expansion now prevail \cite{SNIA,newSN,data}, however we do not yet possess a compelling explanation for what causes the acceleration. Two major efforts for the acceleration mechanism have been 
adding a new substance (termed as dark energy) or reformulating gravity (modifying general relativity) \cite{Carroll:2004de,Caldwell:2009ix,Nojiri:2010wj,Trodden:2011xa}.
By now, a well-established procedure to build and test models in both approaches is first to reproduce the observed redshift-distance relation and second examine the evolution of perturbations for a fixed background expansion. The first step should be a common goal for any dark energy or modified gravity models, and the second can distinguish them from each other 
\cite{Baker:2012zs,Huterer:2013xky}.       

Previously we performed the second step \cite{us-2012, us-2013} on a modified gravity model proposed by Deser and Woodard \cite{DW-2007, DW-2013} once its background was fitted to the expansion history of $\Lambda$CDM \cite{NO-2007, NO-2008, Koivisto, DW-2009}. 
The conclusion was the growth of perturbations predicted by the model is enhanced compared to the one by the 
$\Lambda$CDM model which is statistically disfavored by observations 
\cite{us-2013}.
In fact, this problem of enhanced growth is shared by many modified gravity models including the nonlocal model we considered \cite{reference-for-this}. 

In the current work we challenge this problem by re-adjusting the model with a background expansion rather than that of $\Lambda$CDM. For example, one can find a set of parameters - $\Omega_m$ and a nontrivial equation of state $w$ for dark energy which leads to the same luminosity distance as pointed out in \cite{Shafieloo-Linder2011}. 
Also, it seems to be more natural not to fix the background expansion to $\Lambda$CDM if we want to go beyond $\Lambda$CDM. (We have already checked that this nonlocal model cannot do better than $\Lambda$CDM in suppressing  growth when its background expansion is fixed by $\Lambda$CDM.)

The upshot is the growth rate for this nonlocal model dramatically changes according to the choices of the background expansion: we can find a set of $\{\Omega_m, w \}$ which significantly lowers the growth rate for this nonlocal model. In contrast, for the case of GR the growth rate is not significantly affected by the changes of the background.

\section{Reconstruction of the Expansion}

The model introduces the {\it nonlocal distortion function} $f$ which multiplies the Ricci scalar, to the Einstein-Hilbert Lagrangian \cite{DW-2007},
\begin{equation}
\mathcal{L} = \mathcal{L}_{\rm EH} + \Delta\mathcal{L} =
\frac{1}{16\pi G}\sqrt{-g}\bigg[R + f(X) R
\bigg] \;,
 \eql{action}
\end{equation}
where the argument $X$ of the function $f$ is the inverse scalar d'Alembertian acting on the Ricci scalar, i.e., $X = \square^{-1}R$. One may interpret the function $f$ as a coefficient in front of $R$ which nontrivially modulates the curvature  and henceforth changes the geometry. 

The causal and conserved field equations are derived by varying the action and imposing the retarded boundary conditions on the propagator $\square^{-1}$,
\begin{equation}
G_{\mu\nu} + \Delta G_{\mu\nu} = 8\pi G T_{\mu\nu} \;,
\eql{nonlocal_field_eq}
\end{equation}
where the nonlocal correction to the Einstein tensor takes the form \cite{DW-2007},
\begin{eqnarray}
\lefteqn{
\Delta G_{\mu\nu} 
=
\Bigl[ G_{\mu\nu} \!+\! g_{\mu\nu}\square \!-\! D_{\mu}D_{\nu} \Bigr]
\biggl\{\! f(X)  \!+\! \frac{1}{\square}\Bigl[R f'(X)\Bigr] \!\biggr\}
} \nn \\
&& \hspace {1cm} +
\Bigl[ \delta_{\mu}^{(\rho}\delta_{\nu}^{\sigma)} 
\!-\! \frac{1}{2}g_{\mu\nu}g^{\rho\sigma}\Bigl] 
\partial_{\rho} X
\partial_{\sigma}
\frac{1}{\square}\Bigl[R f'(X)\Bigr]
\;. \quad \quad 
\eql{DeltaGmn}
\end{eqnarray}

The functional form of $f$ can be determined so as to fit a given background geometry. The problem of adjusting $f$, or any parameters of a model in general, to a given geometry is termed as the \textit{reconstruction problem} and the generic procedure for the reconstruction was given in \cite{DW-2009}. In summary, $f$ can be reconstructed by applying the field equations \Ec{nonlocal_field_eq} to the FLRW geometry,
\be
ds^2 = -dt^2 +a^2(t) d\vec{x}\cdot d\vec{x}
\ee
and supposing the scale factor $a(t)$ is known as a function of time.
In the Ref. \cite{DW-2009}, 
$f$ was solved for the case of $\Lambda$CDM expansion.
In the present work, we reconstruct $f$ for the various cases of non-$\Lambda$CDM. Following the notations of  \cite{DW-2009}, we use the dimensionless Hubble parameter $h(\zeta)$ to represent the expansion history and express $f$ in terms of it, 
\begin{eqnarray} 
\lefteqn{f \Bigl(X(\zeta)\Bigr) = -2\int_{\zeta}^{\infty} \!\! d\zetaa \,\zetaa \varPhi(\zetaa)
}
\nn \\
&& \hspace{1cm} - 6 \Omega_{\Lambda} \int_{\zeta}^{\infty} \!\! d\zetaa \, \frac{\zetaa^{
2}}{h(\zetaa) I(\zetaa)} \int_{\zetaa}^{\infty} \!\! d\zetaaa \,
\frac{I(\zetaaa)}{\zetaaa^{ 4} h(\zetaaa)} 
\nonumber\\
& & \hspace{0.8cm} + 2 \int_{\zeta}^{\infty} \!\! d\zetaa \, \frac{\zetaa^{
2}}{h(\zetaa) I(\zetaa)} \int_{\zetaa}^{\infty} \!\! d\zetaaa \,
\frac{r(\zetaaa)\varPhi(\zetaaa)}{\zetaaa^{5}} \;. \quad
\eql{ffin}
\end{eqnarray}
Here, the time variable $\zeta$ is defined in terms of the redshift $z$ as\begin{equation} \nonumber
\zeta \equiv 1 + z = \frac1{a(t)},
\end{equation}
and the dimensionless Hubble parameter $h$ and the dimensionless Ricci scalar $r$ are
\be
h \equiv \frac{H}{H_0}, \quad H \equiv \frac{\dot{a}}{a} 
\quad \mbox{and} \quad 
r \equiv \frac{R}{H_0^2} = 6(\dot{h} + 2h^2)  
\ee
where an overdot denotes a derivative with respect to the cosmic time $t$ and the $H_0$ is the current value of the Hubble parameter. The functions $\varPhi(\zeta)$ and $I(\zeta)$ are given by \cite{DW-2009},
\bea
\varPhi(\zeta) &=& -6\Omega_{\Lambda} \int_{\zeta}^{\infty} \!\! d\zetaa \,
\frac1{h(\zetaa)} \int_{\zetaa}^{\infty} \!\! d\zetaaa \, 
\frac1{\zetaaa^{4} h(\zetaaa)} \;,
\\
I(\zeta) &=& \int^\infty_\zeta d\zetaa \frac{r(\zetaa)}{\zetaa^{4}h(\zetaa)}\; .
\eql{Phi_Idef}
\eea
We take the expression of $h(z)$ employed in \cite{Shafieloo-Linder2011} as 
a non-$\Lambda$CDM expansion\footnote{We ignore the spatial curvature included in $h(z)$, thus in our expression
$\Omega_{\Lambda} \approx  \Omega_{de} \approx 1-\Omega_m$.},
\be
h^2(\zeta) = \Omega_{m}\zeta^3 + \Omega_{de}\exp\biggl[3 \! \int_{1}^{\zeta} \!d \zeta' \frac{1+w(\zeta')}{\zeta'} \biggr] \;,
\eql{h_non-lcdm}
\ee   
and numerically integrate\ \Ec{ffin} to get the distortion function $f$.  
Note that for the case of $\Lambda$CDM expansion, $h^2(\zeta) = \Omega_{\Lambda} + \Omega_m \zeta^3 + \Omega_r \zeta^4$, all the expressions in \Ec{ffin} and \Ec{Phi_Idef} recover their forms in \cite{DW-2009}.

The point is that once the reconstruction of $f$ is done, the model automatically fulfills the first goal of reproducing 
a given expansion history. 
The next step is to examine the growth of perturbations with different $f$'s corresponding to the various (non-$\Lambda$CDM) expansions determined by the free parameters $\Omega_m$ and $w$.  
Good news is that we can find the parameter sets of $\{\Omega_m, w\}$ and relatively suppressed growth rate. The result is presented in the next section.

\section{Growth of Perturbations} 

We perturb the metric around the FLRW background as
\be
ds^2 = -\left(1\!+\!2\Psi(t,\vec x)\right) dt^2 + a^2(t) dx^2 \left(1\!+\!2\Phi(t,\vec x)\right)\eql{metric} .
\ee
By substituting the perturbed metric back in the nonlocal field equation \Ec{nonlocal_field_eq} and expanding it to the first order, we obtain the evolution equations for the perturbations \cite{us-2012},
\bea
\lefteqn{
(\Phi + \Psi) = -(\Phi + \Psi)\biggl\{f(\overline{X}) +
\frac{1}{\overline{\square}}\bigg[\overline{R}f'\Bigl(\overline{X}\Bigr)\bigg]\biggl\}
}
\nn \\
&& \hspace{1.3cm} -\biggl\{ f'(\overline{X})\frac{1}{\overline{\square}}\delta R +
\frac{1}{\overline{\square}}
\bigg[f'\Bigl(\overline{X}\Bigr)\delta R \bigg]\biggr\}\;, \quad
\eql{aniso}
\\
\lefteqn{
\frac{k^2}{a^2}\Phi 
+\frac{k^2}{a^2}\Biggl[
\Phi\biggl\{f(\overline{X}) 
+ \frac{1}{\overline{\square}}\bigg[\overline{R}f'\Bigl(\overline{X}\Bigr)\bigg] \biggr\} 
}
\nn \\
&& \hspace{-0.2cm} + \frac{1}{2} 
\Biggl\{
f'(\overline{X})
\frac{1}{\overline{\square}}\delta R + 
\frac{1}{\overline{\square}} \left[ f'\Bigl(\overline{X}\Bigr)\delta R \right]
 \biggr\}
 \Biggr]
= 4\pi G\bar{\rho} \delta \;. \quad
\eql{modpos}
\eea
Here $\bar{\rho}$ is the mean matter density and $\delta$ is the fractional over-density in matter.
These two are compared with the corresponding perturbation equations in general relativity (GR),\bea
(\Phi + \Psi)&=& 0\;,
\eql{anisoGR}
\\
\frac{k^2}{a^2}\Phi &=& 4\pi G\bar{\rho} \delta \;.
\eql{posGR}
\eea
Since the modified field equations \Ec{nonlocal_field_eq} are conserved, i.e., $\nabla^\mu \Delta G_{\mu\nu} = 0$,
the two conservation equations in GR still hold in the nonlocal model, 
\bea
\dot{\delta}+H\theta &=& 0 \;,
\eql{delta}
\\ 
H\dot{\theta} + \Bigl(\dot{H} + 2 H^2 \Bigr)\theta -\frac{k^2}{a^2}\Psi &=& 0 \;,
\eql{theta}
\eea
and complete the system of the evolution equations for the four perturbation variables, $\Phi, \Psi, \delta$ and 
$\theta \equiv \nabla \cdot \vec{v}/H$. Here $\vec{v}$ is the comoving peculiar velocity.
Combining the four equations leads to the equation governing the growth of perturbations,
\be
\frac{d^2\delta}{d\zeta^2} 
+ \biggl[ \frac{1}{h(\zeta)}\frac{d h(\zeta)}{d\zeta} - \frac{1}{\zeta} \biggr]\frac{d\delta}{d\zeta} 
- \frac{3}{2}(1+\mu)\Omega_m \frac{\zeta}{h^2(\zeta)}\delta = 0
\eql{delta_eqn_zeta}
\ee
Note that the deviations from GR in the nonlocal model are encoded into the parameter $\mu$ devised in \cite{us-2013:9and19}. Hence, when the background is fixed by the one of GR, i.e., the $\Lambda$CDM
expansion, the only factor differs from GR in this equation \Ec{delta_eqn_zeta} is $\mu$. 
The growth of $\delta$ is then simply determined by the sign of $\mu$: positive (negative) $\mu$ gives enhanced (suppressed) growth. In our previous analysis for this model, it turned out to be positive and hence we concluded that growth is enhanced in the nonlocal model \cite{us-2013}. 

Now, we note that the effect of non-$\Lambda$CDM backgrounds for which the Hubble expansion rate $h$ is different from the one for $\Lambda$CDM - denote it $h_{\Lambda}$. The non-$\Lambda$CDM expansion rate changes both the source term (not only $\mu$ but also $h^2$ in the denominator) and the friction term in the growth equation \Ec{delta_eqn_zeta}, and hence leads to more complicated dynamics. Before analyzing it, let us emphasize that the conclusion we made in \cite{us-2013} was based on fixing the background with $\Lambda$CDM and the growth equation we used was actually
\be
\frac{d^2\delta}{d\zeta^2} 
+ \biggl[ \frac{1}{h_{\Lambda}(\zeta)}\frac{d h_{\Lambda}(\zeta)}{d\zeta} - \frac{1}{\zeta} \biggr]\frac{d\delta}{d\zeta} 
- \frac{3}{2}\Bigl[1+\mu(h_\Lambda)\Bigr]\Omega_m \frac{\zeta}{h_{\Lambda}^2(\zeta)}\delta = 0
\eql{delta_eqn_zeta_hLambda}
\ee
Here $\mu$ is a function of $f$ which is determined by $h$, hence it is eventually a function of $h$.
 
\subsection{$\Lambda$CDM vs. Non-$\Lambda$CDM Backgrounds}

\begin{figure*}[!t]
\begin{center}
 \begin{tabular}{ccc} 
  \includegraphics[width=0.31\textwidth]{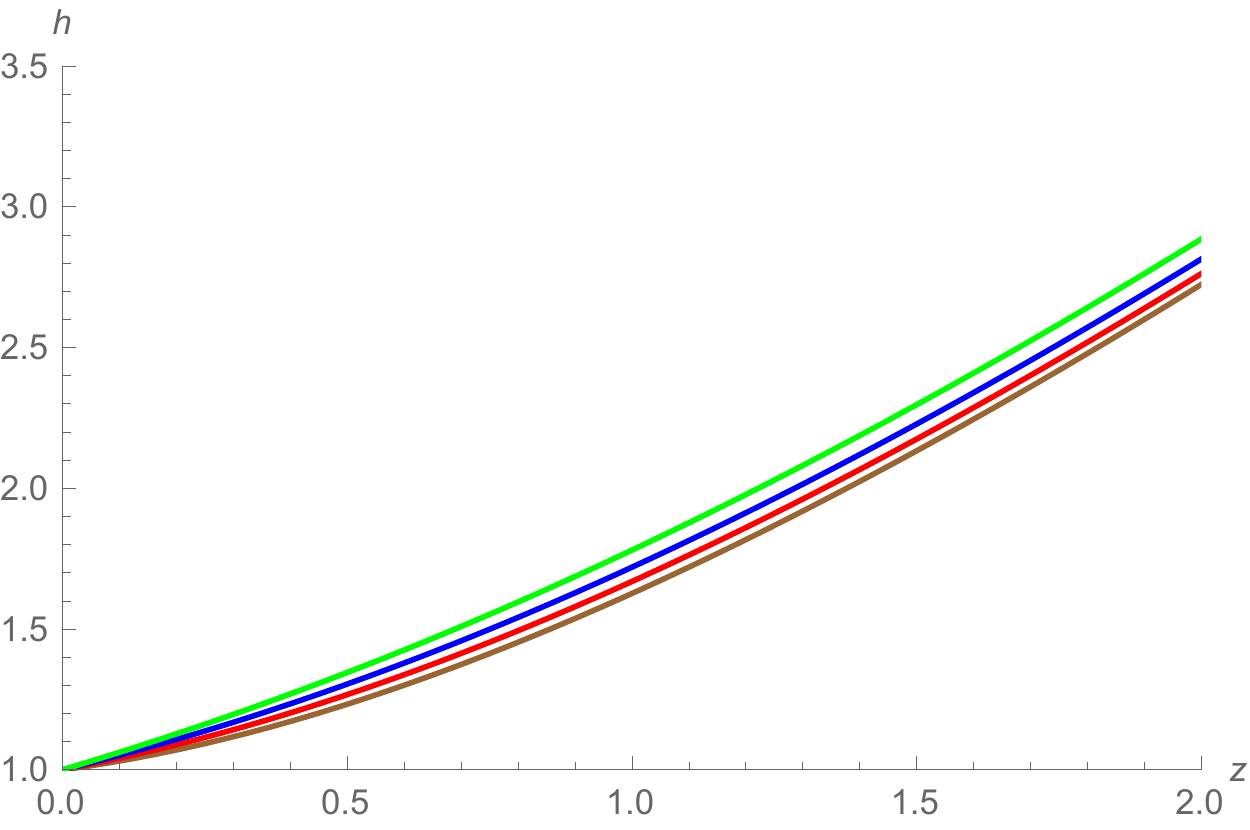}
  &
  \includegraphics[width=0.31\textwidth]{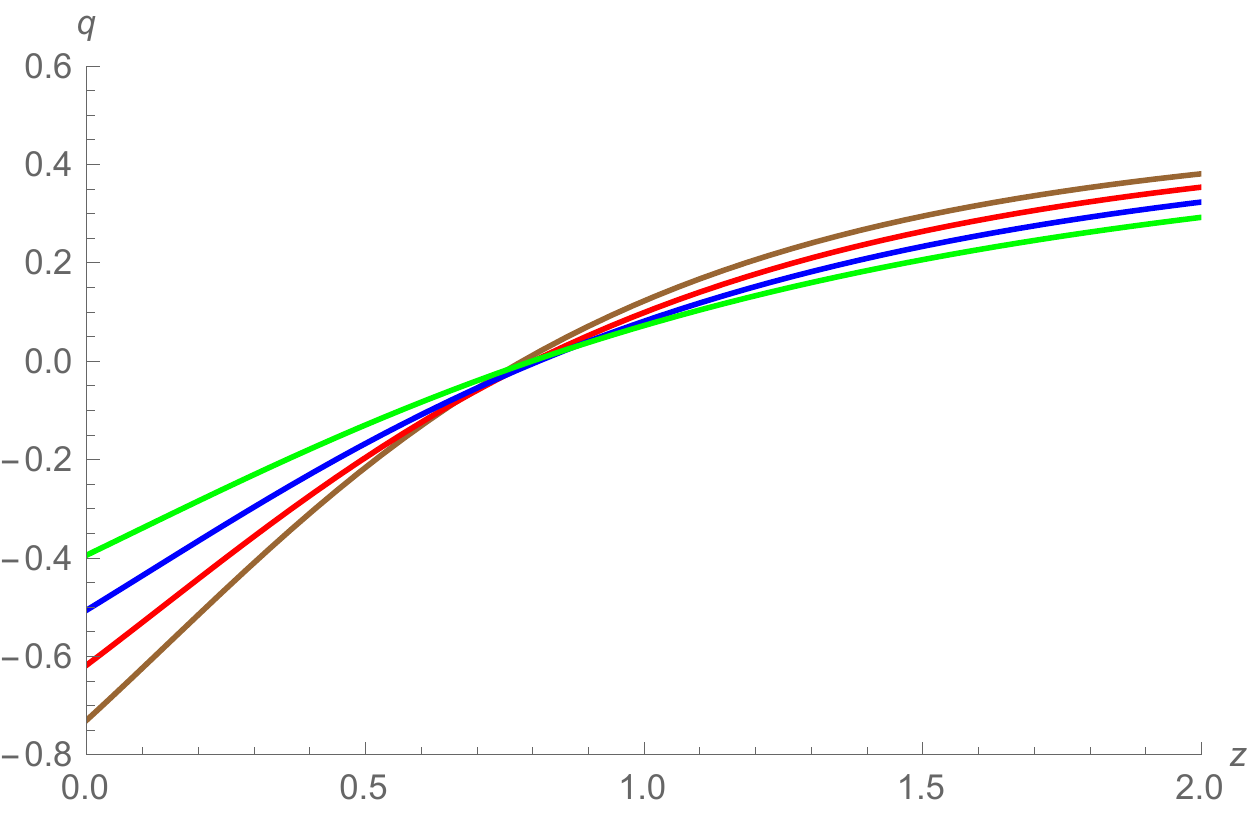}
  &
  \includegraphics[width=0.31\textwidth]{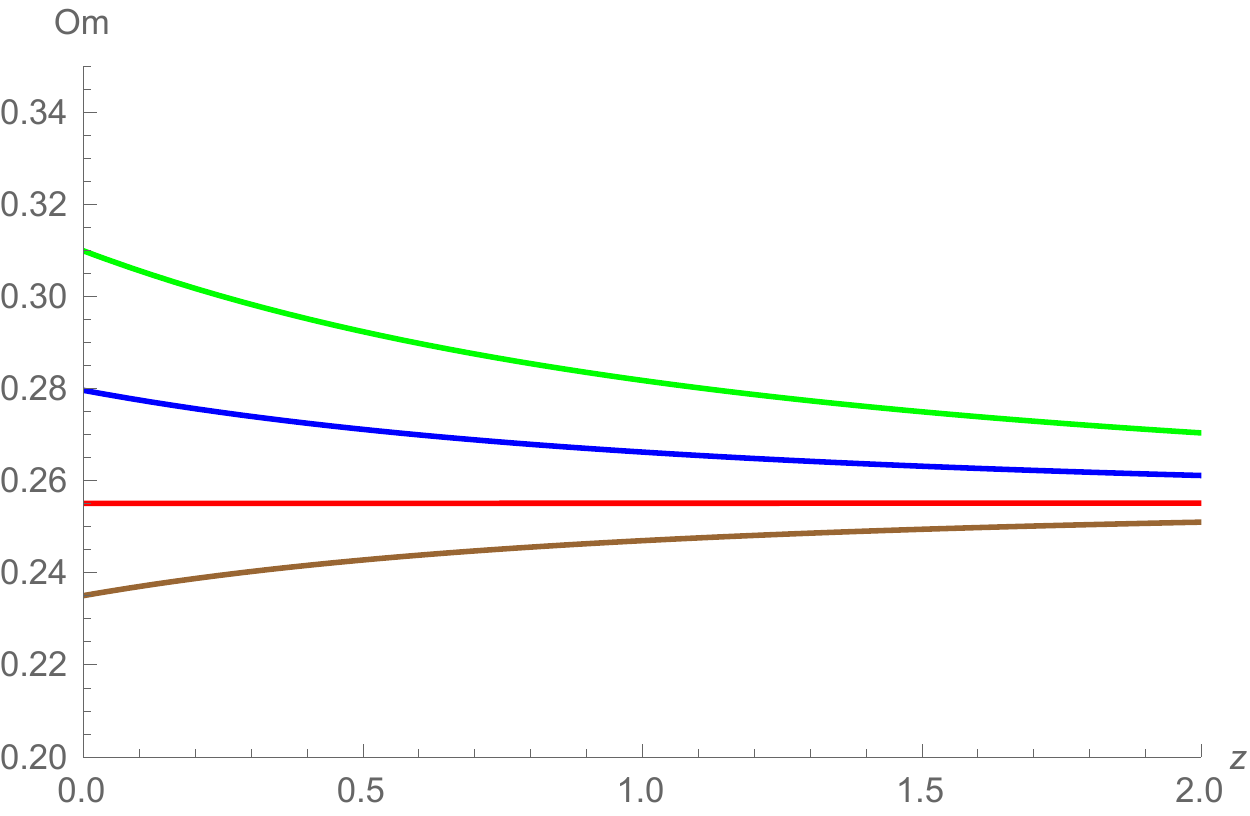} \end{tabular}
\end{center}
\caption{The dimensionless Hubble parameter, $h(z)$, the deceleration parameter, $q(z)$ and
the \textit{Om} diagnostic, $Om(z)$ as a function of redshift for $\Omega_m = 0.255$ and $w=-0.8$ (green), $-0.9$ (blue), $-1$
(red), $-1.1$ (brown) respectively.} 
\label{fig:h-q-Om}
\end{figure*}

First, we investigate the features of non-$\Lambda$CDM backgrounds. For simplification we focus on the cases where the equation of state $w$ is constant so that the dimensionless Hubble expansion rate \Ec{h_non-lcdm} becomes
\be
h^2(\zeta) = \Omega_{m}\zeta^3 + (1-\Omega_m)\zeta^{3(1+w)} \;,
\eql{h_non-lcdm_const-w}
\ee     
where we set $\Omega_{de} = 1- \Omega_m$. We survey the three quantities - the dimensionless Hubble parameter,   
the deceleration parameter and the \textit{Om} diagnostic - for different values of parameters $\Omega_m$ and $w$. 
For example, Fig. \ref{fig:h-q-Om} depicts these three for the fixed value of $\Omega_m = 0.255$ and different $w$'s. 

The dimensionless Hubble parameter increases as $w$ becomes less negative, as easily expected from \Ec{h_non-lcdm_const-w}, which would mean the source term in \Ec{delta_eqn_zeta} gets smaller ($h^2$ being in the denominator) and leads to suppressed growth.

The deceleration parameter $q(\zeta) = -1 +\frac{\zeta h'(\zeta)}{h(\zeta)}$, where prime means a derivative with respect to $\zeta$, affects the friction term in \Ec{delta_eqn_zeta}. It changes more dramatically for more negative $w$ so that the absolute value of $q$ is larger in the high (or deceleration phase $q > 0$) and low (or acceleration phase $q<0$) redshift, which would lead to more friction and suppressed growth. 
But we will see later the reduction in the source term is more important 
so less negative $w$ actually better suppress the growth.

The \textit{Om} diagnostic defined by ~\cite{om,om_chris} 
\be
Om(z) \equiv \frac{h^2(z) - 1}{(1+z)^3 -1}
\ee 
measures how much $h$ deviates from the $\Lambda$CDM value. 
The farther it is from the horizontal straight line of $\Lambda$CDM, 
the more different from $\Lambda$CDM.
Of course these background effects enter both GR and modified gravity models including this model. 
However, the changes in the background have more influence on this nonlocal model than on GR as we will see shortly.

\subsection{The Growth Rate}

\begin{figure}[htbp]
\centering
\includegraphics[width=\columnwidth]{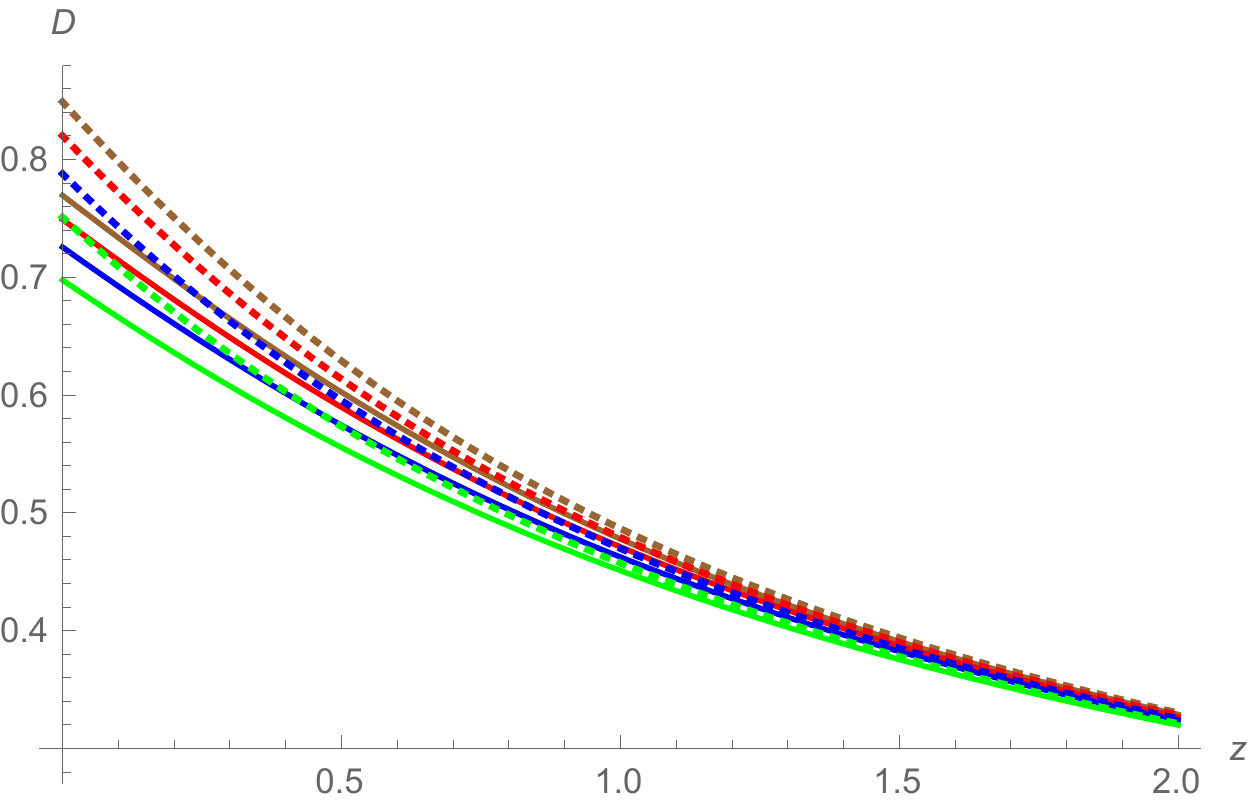}
\caption{The growth function, $D(z)$ as a function of redshift in GR (solid lines) and the nonlocal (dotted lines) model $\Omega_m = 0.255$ and $w=-0.8$ (green), $-0.9$ (blue), $-1$ 
(red), $-1.1$ (brown) respectively.} 
\label{fig:D-Om0255}
\end{figure}

The growth function, $D(\zeta)$, is the solution to \Ec{delta_eqn_zeta} with initial conditions $D(\zeta) = 1/\zeta$ at early times when matter still dominates ($z \simeq 10$). Fig. \ref{fig:D-Om0255} depicts the growth function in GR and the nonlocal model with different background expansions corresponding to Fig. \ref{fig:h-q-Om}. 
The initial condition at $z_{\rm init} = 9$ was set the same for each solution. As noted above, less negative $w$ makes the source term in \Ec{delta_eqn_zeta} smaller and hence lowers the growth for both GR and the nonlocal model.

\begin{figure*}[!t]
\begin{center}
 \begin{tabular}{ccc} 
  \includegraphics[width=0.42\textwidth]{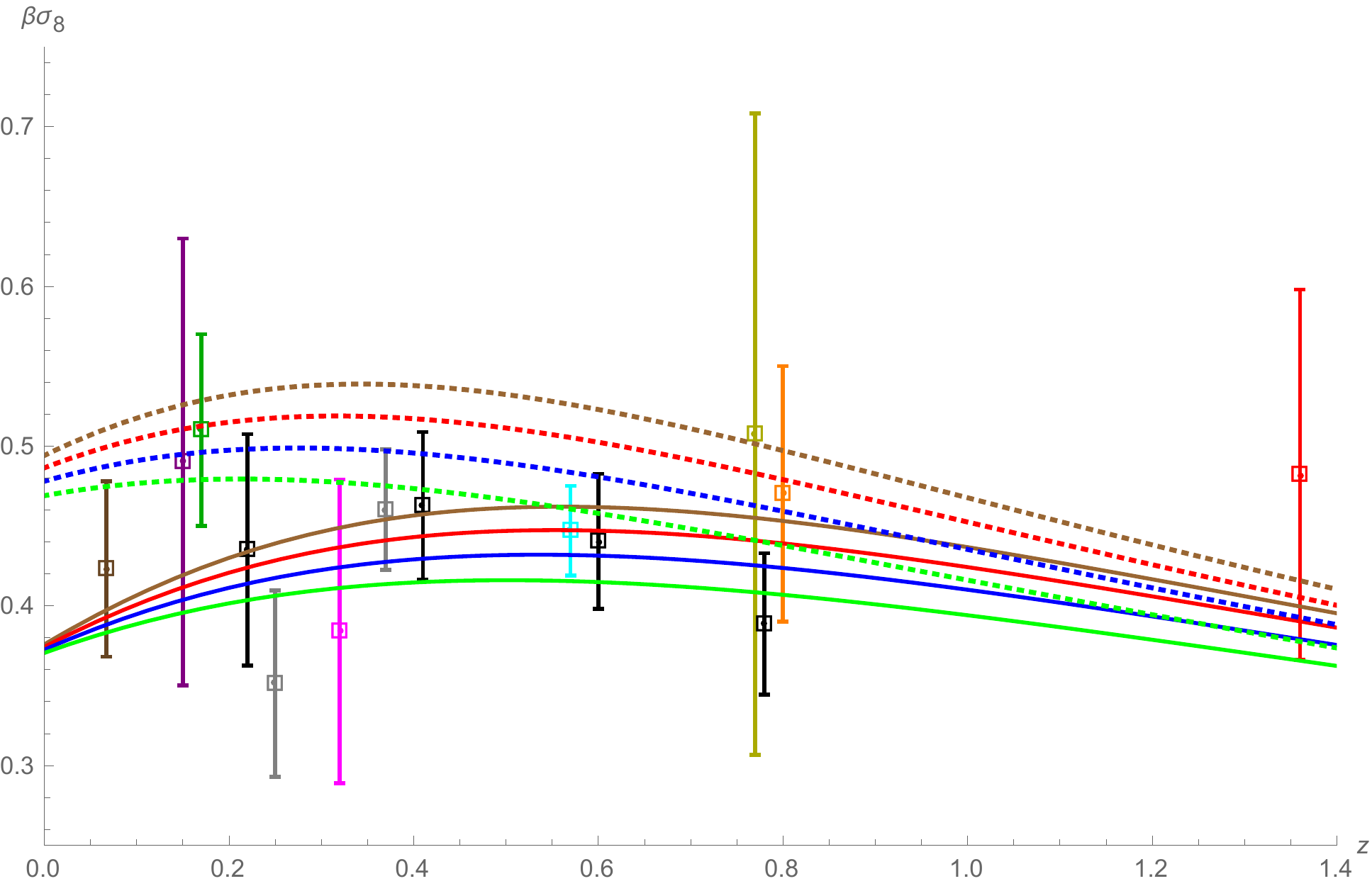}
  &
  \includegraphics[width=0.42\textwidth]{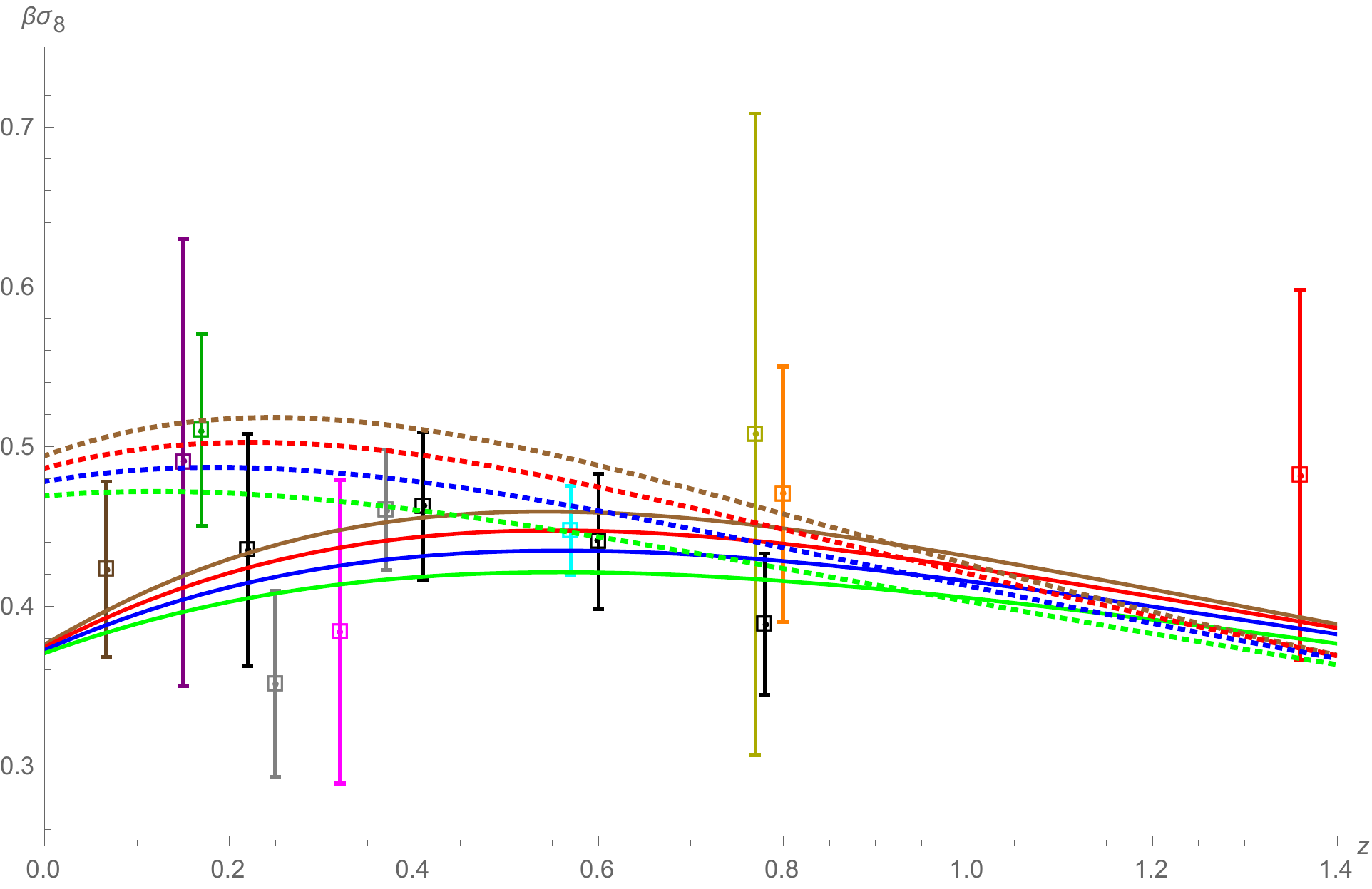}
  &
  \includegraphics[width=0.1\textwidth, height=4.8cm]{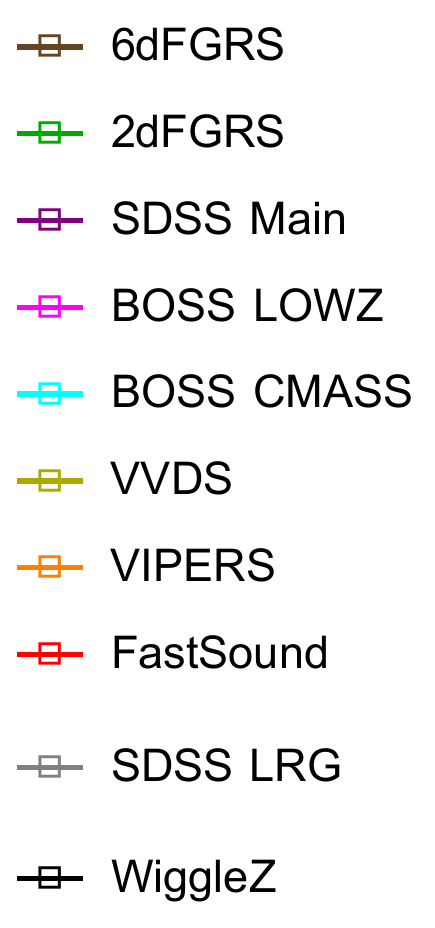} 
 \end{tabular}
\end{center}
\caption{The growth rate  $\beta(z)\sigma_8(z)$ as a function of redshift in GR (solid lines) and the nonlocal model (dotted lines) for $\Omega_m = 0.255$ and$w=-0.8$ (green), $-0.9$ (blue), $-1$ 
(red), $-1.1$ (brown) respectively. At the left panel $\sigma_8(z=9)$ is set the same and at the right panel $\sigma_8(z = 0)$ the same for each solution of the growth equation \Ec{delta_eqn_zeta}. The fluctuation amplitude today was chosen as $\sigma_8 = 0.8$ following \cite{BOSS-2016}. Data points
come from 6dFGRS, 2dFGRS, SDSS main galaxies, SDSS LRG, BOSS LOWZ, WiggleZ, BOSS CMASS, VVDS and VIPERS \cite{Teppei}. (The numbers of the data points are taken from Figure 17 of \cite{Teppei} with the aid of Teppei Okumura.) The most recent BOSS data \cite{BOSS-2016} are not used here, however including them will not change results much.} 
\label{fig:fsigma8}
\end{figure*}

The product of the growth rate 
$\beta\equiv d\ln D/d\ln a$ and the fluctuations amplitude $\sigma_8$ 
is a quantity directly measured in spectroscopic surveys. 
We examine two slightly different normalization conditions for the growth rate:
One way is to set  the initial amplitude $\sigma_8(z_{\rm init})$ the same for GR and the nonlocal using the growth function of GR (with $\Lambda$) and $\sigma_8(z=0)$, 
which is the method also employed in \cite{us-2013}, 
\be
\sigma_8(z_{\rm init}) = \sigma_8(z=0) \,\frac{D^{\rm GR}(z_{\rm init})}{D^{\rm GR}(0)} \;.
\eql{sigma8-set-at-9}
\ee
In this case, the theoretically computed $\sigma_8(z)$ using each solution of the growth equation does not evolve to the measured $\sigma_8(z=0)$. 
The other way is to set the amplitude today  $\sigma_8(z=0)$ the same 
using their own growth functions $D(z)$ (see the 8 different growth functions depicted in Fig. \ref{fig:D-Om0255}), 
\be
\sigma_8(z) = \sigma_8(z=0) \,\frac{D(z)}{D(0)} \;.
\eql{sigma8-set-at-0}
\ee
In this case, $\sigma_8(z=0)$ is guaranteed to be the same but $\sigma_8(z_{\rm init})$ computed by \Ec{sigma8-set-at-0} is different for each solution.
Fig. \ref{fig:fsigma8} shows both cases: at the left panel  $\sigma_8(z_{\rm init})$ is set the same and at the right panel $\sigma_8(z = 0)$ the same for each solution. 
In the sense of fitting the growth data, the nonlocal model with the background of a slightly less negative equation of state  does a remarkable job with this second normalization condition: $\chi^2 = 7.88$ for GR with $w=-1$ \textit{vs.} $8.44$ for the nonlocal model with $w=-0.8$.  The $\chi^2$ values for these eight solutions with the two normalization conditions are summarized in the Table \ref{tab:chi^2}.

\begin{table*}[!t]
\begin{center}
\caption{The $\chi^2$ values between the 
data points and the $\beta(z)\sigma_8(z)$ predicted by GR and the nonlocal model with four different background expansion histories. The same $\sigma_8(z=9)$ means it is normalized by \Ec{sigma8-set-at-9} and the same $\sigma_8(z=0)$ by \Ec{sigma8-set-at-0}. }
\label{tab:chi^2}

\begin{tabular}{c|cccc|cccc}
\hline
\hline
 & \multicolumn{4}{c|}{GR} &  \multicolumn{4}{c}{Nonlocal} \\
Same  & $~~w=-1.1~~$ & $~~w=-1~~$  & $~~w=-0.9~~$ &$~~w=-0.8~~$ & $~~w=-1.1~~$ & $~~w=-1~~$  & $~~w=-0.9~~$ &$~~w=-0.8~~$  \\
\cline{2-9}
$\sigma_8(z=9)$ & $8.69$ & $7.88$ & $8.85$ & $11.91$ & $42.75$ &$28.46$  & $17.45$ & 10.34\\
$\sigma_8(z=0)$ & $8.35$ & $7.88$ & $8.60$ & $10.75$ & $24.29$ &$17.13$  & $11.70$ & $8.44$\\
\hline
\end{tabular}

\end{center}
\end{table*}

\section{Discussion}\label{discuss}

We have analyzed the growth of perturbations predicted by a nonlocal gravity model of type \Ec{action}.
The nonlocal distortion function $f$ can be constructed to reproduce any desired expansion history. (As remarked in \cite{DW-2009}, ``absent a derivation from fundamental theory, $f$ has the same status as the potential $V(\varphi)$ in a scalar quintessence model and the function $F(R)$ in $F(R)$ gravity''.)
Once the function $f$ is fixed, no free parameter is remained and hence the evolution of perturbations is fully governed by the model's rule for gravity and the expansion history it chose to mimic.  
That is, the growth of perturbations depends on the background expansion which they ride on. 
Previously we have found that when the background is chosen to be the exact $\Lambda$CDM,
the model enhances growth compared to that predicted in general relativity with $\Lambda$, which is disfavored by measurements \cite{us-2013}. 

In the present paper, we have examined the background effects and found that the model can suppress growth when its background is chosen to be some particular non-$\Lambda$CDM expansion.  
A non-$\Lambda$CDM background with the equation of state for dark energy $w$ less negative than $-1$ tends to lower the growth. Notably, this tendency is more dramatic in the nonlocal model than in GR. That is, the statistical significance substantially improves for this nonlocal model with the slight change of $w$ whereas it does not vary much for GR for the same change of $w$. (see Table \ref{tab:chi^2}).
While the growth rate data still appear best fit to GR with $w=-1$ ($\chi^2 = 7.88$), the statistical difference from the nonlocal model with $w=-0.8$ ($\chi^2 = 8.44$ or $10.34$ depending on the normalization conditions) is insignificant. 
In summary, relaxing the condition of the background being exactly $\Lambda$CDM and setting $w$ less negative than $-1$ tends to lower the growth of perturbations; however, further examination of observationally allowed forms of $w(z)$ is needed. 

The interpretation of the tight parametric constraints on the equation of state of dark energy using CMB data should be performed carefully. In general, CMB data alone are not very suitable to directly study dark energy, hence they are usually used in combination with other data assuming a parametric form for the expansion history. Constraints obtained from such model fitting analysis can be sometimes conflictive (although they may hint towards some new physics in the data). 
Therefore, conclusions cannot be easily drawn without extensive analysis and support from different observations. For instance, Planck CMB data has already shown some conflicts with other cosmology surveys in estimation of the value of Hubble constant $H_0$ (assuming concordance $\Lambda$CDM model) \cite{Planck2015-13}. 
Another recent major survey analysis pointed out some discrepancy between $H_0$ and Lyman-$\alpha$ forest BAO data when assuming $\Lambda$CDM model \cite{Zhao-2017}. Hence, we have considered the direct distance-scale data such as standard candles and rulers for the expansion history rather than using constraints obtained by model fitting analysis of CMB data. This is particularly important in the present work, since we are discussing a very different cosmology model which in turn generates different perturbations. It should also be noted that the most recent compilation of supernovae data known as JLA compilation \cite{SDSS-2014} does not rule out $w=[-1.1, -0.8]$ for the constant equation of state of dark energy.

Our next step is to look for a choice of the background expansion history extensively (not just for constant $w$ but more nontrivial evolution of it),  within the flexibility data allows, that leads to a reasonable fit to the growth data with this nonlocal model.

Another purpose of this work is to recall the usage of nonlocal modifications for gravity and aid model builders to extend it. The nonlocal invariant $X=\square^{-1}R$ is the simplest, so easy to handle with, that's why we chose to analyze it first. However, as pointed out by Woodard (one of the inventors of this model) \cite{W-review-nonlocal-2014}, it achieves acceleration by strengthening gravity which would lead to enhanced growth. (That's why less negative $w$ meaning less acceleration works better for it.) A better way would be to make a nonlocal model emulates time-varying cosmological constant \cite{W-review-nonlocal-2014}. There have been projects of building nonlocal gravity model in this direction, but have not reached the level to fully describe the phenomenology of the late-time cosmic acceleration. Those extended models inherit the main virtues of this simplest, $f(\square^{-1}R)$ model: 
\begin{itemize}
\item{$\square^{-1}R$ is a dimensionless quantity so that no new mass parameter is introduced.} 
\item{$\square^{-1}R$ grows slowly so it does not require huge fine-tuning.} 
\item{It evades any deviation from general relativity for the solar system by exploiting the sign of $\square^{-1}R$}
\item{It does not require an elaborate screening mechanism to avoid kinetic instability (or ghosts.)}
\end{itemize}

It is also worth noting that the distinction of this class of models from the nonlocal gravity models proposed and developed by Maggiore and his collaborators \cite{Maggiore}. In those models, nonlocal invariants are not dimensionless and multiplied by a free parameter of $mass^2$ dimension. The mass, whose origin seems not to be known yet, plays a role of cosmological constant and no arbitrary function (like the nonlocal distortion function $f$) exists in those models. 
The phenomenology of those nonlocal models has been reported to be very successful and investigation for deriving those nonlocalities from a fundamental theory through quantum processes is in progress \cite{Maggiore, Akrami}. 
It would be interesting to compare the approaches to nonlocal gravity by these two groups and relate their origins from fundamental theories. 

\textit{Additional note:} A very recent paper by Nersisyan, Cid and Amendola \cite{Amendola2017} claims that a localized version of this nonlocal model when its background expansion is fixed by $\Lambda$CDM leads to suppressed growth for the perturbations, which is opposite to our previous result \cite{us-2013}. If their analysis turns out correct, localizing could be a better way to have lower growth than changing the background (as studied in the present paper). However, we find that their implementation of the sub-horizon limit is different from ours \cite{us-2012,us-2013} and it is the main source of the discrepancy.  Currently we are collaborating with them to further investigate this issue and we will jointly report the detailed explanations for it.  

\vskip 1cm

\centerline{\bf Acknowledgements}
We thank Scott Dodelson, Emre Kahya, Seokcheon Lee, Eric Linder, Chris Sabiu, Richard Woodard and Yi Zheng for useful suggestions and conversations. We acknowledge and thank Teppei Okumura for providing us with the growth rate data points and Scott Dodelson and Richard Woodard for reading our manuscript.


\begin{thebibliography}{99}

\bibitem{SNIA} A. G. Riess, {\it et al}., Astron. J. {\bf 116} (1998) 1009,
astro-ph/9805201; S. Perlmutter, {\it et al}., Astrophys. J. {\bf 517}
(1999) 565, astro-ph/9812133.

\bibitem{newSN} M. Hicken et al., Astrophys. J. {\bf 700} (2009)
1097, arXiv:0901.4804.

\bibitem{data} WMAP Collabration, Astrophys. J. Suppl. {\bf 208} (2013) 19, arXiv:1212.5226; 
Z. Hou et al., Astrophys. J. {\bf 782} (2014) 74,  arXiv:1212.6267; 
J. L. Sievers {\it et al}., JCAP {\bf 1310} (2013) 060, arXiv:1301.0824; 
Planck Collaboration, Astron. Astrophys. {\bf 571} (2014) A16, arxiv:1303.5076;
Planck Collaboration, Astron. Astrophys. {\bf 594} (2016) A24, arXiv:1502.01597.

\bibitem{Carroll:2004de} S. M. Carroll, A. De Felice, V. Duvvuri, D. A. Easson, M. Trodden, {\it et al}., Phys. Rev. {\bf D71} (2005) 063513, arXiv:astro- ph/0410031.
\bibitem{Caldwell:2009ix} R. R. Caldwell and M. Kamionkowski, Ann. Rev. Nucl. Part. Sci. {\bf 59} (2009) 397, arXiv:0903.0866.
\bibitem{Nojiri:2010wj} S. Nojiri and S. D. Odintsov, Phys. Rept. {\bf 505}  (2011) 59, arXiv:1011.0544.
\bibitem{Trodden:2011xa} M. Trodden, Gen. Rel. Grav. {\bf 43} (2011) 3367, arXiv:1105.0721.

\bibitem{Baker:2012zs} T. Baker, P. G. Ferreira, and C. Skordis, Phys. Rev. {\bf D87} (2013)
024015, arXiv:1209.2117.

\bibitem{Huterer:2013xky} D. Huterer, D. Kirkby, R. Bean, A. Connolly, K. Dawson,
et al.,  Astropart. Phys. {\bf 63} (2015) 23, arXiv:1309.5385.

\bibitem{us-2012} S. Park and S. Dodelson, Phys. Rev. {\bf D87} (2013) 024003, arXiv:1209.0836.
\bibitem{us-2013} S. Dodelson and S. Park,  Phys. Rev. {\bf D90} (2014) 043535, arXiv:1310.4329.

\bibitem{DW-2007} S. Deser and R. P. Woodard, Phys. Rev. Lett. {\bf 99} (2007), arXiv:0706.2151.
\bibitem{DW-2013} S. Deser and R. P. Woodard, JCAP {\bf 11} (2013) 036, arXiv:1307.6639.

\bibitem{NO-2007}  
S. Nojiri and S. D. Odintsov, Phys. Lett. {\bf B659} (2008) 821, arXiv:0708.0924.
\bibitem{NO-2008} 
S. Jhingan, S. Nojiri, S. D. Odintsov, M. Sami, I. Thongkool, S. Zerbini, Phys. Lett. {\bf B663} (2008) 424, arXiv:0803.2613.
\bibitem{Koivisto} T. Koivisto, Phys. Rev. {\bf D77} (2008) 123513, arXiv:0803.3399.
\bibitem{DW-2009} C. Deffayet and R. P. Woodard, JCAP {\bf 08} (2009) 023, arXiv:0904.0961.

\bibitem{reference-for-this} T. Clifton, P. G. Ferreira, A. Padilla, C. Skordis, Phys. Rept. {\bf 513} (2012) 1, arXiv:1106.2476; K. Koyama, Rept. Prog. Phys. {\bf 79} (2016) 046902, arXiv:1504.04623; 
A. Joyce, L. Lombriser, and F. Schmidt, arXiv:1601.06133.

\bibitem{Shafieloo-Linder2011} A. Shafieloo and E. Linder, Phys. Rev. {\bf D84} (2011) 063519, arXiv:1107.1033.   

\bibitem{us-2013:9and19} F. Simpson, C. Heymans, D. Parkinson, C. Blake, M. Kil-
binger, et al., Mon. Not. Roy. Astron. Soc. \textbf{429} (2013) 2249, arXiv:1212.3339. 
L. Amendola, M. Kunz, and D. Sapone, JCAP \textbf{0804} (2008) 013, 0704.2421.

\bibitem{om}
V. Sahni, A. Shafieloo and A. A. Starobinsky, Phys. Rev. {\bf D78} (2008) 103502.
\bibitem{om_chris}
C. Zunckel and C. Clarkson, Phys. Rev. Lett. {\bf 101} (2008) 181301.

\bibitem{BOSS-2016} S. Alam {\it et al}., BOSS Collaboration, arXiv:1607.03155.

\bibitem{Teppei} T. Okumura {\it et al}., Publ. Astron. Soc. Jap. {\bf 68} (2016) 38, arXiv: 1511.08083. 

\bibitem{W-review-nonlocal-2014}   R. P. Woodard, Found. Phys. {\bf 44} (2014) 213, arXiv:1401.0254

\bibitem{Planck2015-13} P. A. R.  Ade {\it et al}., Planck Collaboration, Astron. Astrophys. {\bf 594} (2016) A13, arXiv:1502.01589.

\bibitem{Zhao-2017} Zhao {\it et al}., arXiv:1701.08165.

\bibitem{SDSS-2014} M, Betoule {\it et al}., SDSS Collaboration, Astron. Astrophys. {\bf 568} (2014) A22, arXiv:1401.4064.

\bibitem{Maggiore} 
M. Maggiore, Phys. Rev. {\bf D89} (2014) 043008, arXiv:1307.3898;
S. Foffa, M. Maggiore, and E. Mitsou, Phys. Lett. {\bf B733} (2014) 76, arXiv:1311.3421;
S. Foffa, M. Maggiore, and E. Mitsou, Int. J. Mod. Phys. {\bf A29} (2014) 1450116, arXiv:1311.3435;
A. Kehagias and M. Maggiore, JHEP {\bf 1408} (2014) 029, arXiv:1401.8289;
M. Maggiore and M. Mancarella, Phys.Rev. {\bf D90} (2014) 023005, arXiv:1402.0448;
Y. Dirian, S. Foffa, N. Khosravi, M. Kunz, and M. Maggiore, JCAP {\bf 1406} (2014) 033, arXiv:1403.6068;
Y. Dirian, S. Foffa, M. Kunz, M. Maggiore, and V. Pettorino, JCAP {\bf 1504} (2015) 044, arXiv:1411.7692;
E. Mitsou, Aspects of Infrared Non-local Modifications of General Relativity, Ph.D. thesis, Geneva U. (2015), arXiv:1504.04050; 
M. Maggiore, arXiv:1506.06217;
G. Cusin, S. Foffa, M. Maggiore, and M. Mancarella, Phys. Rev. {\bf D93} (2016) 043006, arXiv:1512.06373;
G. Cusin, S. Foffa, M. Maggiore, and M. Mancarella, Phys. Rev. {\bf D93} (2016) 083008, arXiv:1602.01078;
Y. Dirian, S. Foffa, M. Kunz, M. Maggiore, and V. Pettorino, JCAP {\bf 1605} (2016) 068,  arXiv:1602.03558;
M. Maggiore, Phys. Rev. {\bf D93} (2016) 063008, arXiv:1603.01515;
M. Maggiore, arXiv:1606.08784.
\bibitem{Akrami} H. Nersisyan, Y. Akrami, L. Amendola, T. S. Koivisto, and J. Rubio, Phys. Rev. {\bf D94} (2016) 043531, arXiv:1606.04349.

\bibitem{Amendola2017} H. Nersisyan, A. F. Cid, and L. Amendola, arXiv:1701.00434. 

\end{thebibliography}
\end{document}